\documentclass[prl,reprint,amsmath,amssymb,aps]{revtex4-1}
\pdfoutput=1
\usepackage{verbatim}
\usepackage{fullpage}
\usepackage{graphicx}
\usepackage{graphicx}
\usepackage{dcolumn}
\usepackage{bm}
\newcommand{\be}{\begin{equation}}
\newcommand{\ee}{\end{equation}}
\newcommand{\bea}{\begin{eqnarray}}
\newcommand{\eea}{\end{eqnarray}}

\newcommand{\bnabla}{\mbox{\boldmath $\nabla$}}
\renewcommand{\epsilon}{\varepsilon}

\begin{document}
\title{
Skyrmion knots in frustrated magnets
}
\author{Paul Sutcliffe\\ \ }
\affiliation{
Department of Mathematical Sciences,
Durham University, Durham DH1 3LE, United Kingdom.\\ 
Email: p.m.sutcliffe@durham.ac.uk}
\date{June 2017}

\begin{abstract}
  A magnetic Skyrmion is a stable two-dimensional nanoparticle
  describing a localized winding of the magnetization in certain
magnetic materials.
  Skyrmions are the subject of
  intense experimental and theoretical investigation and have potential technological spintronic applications.
  Here we show that numerical computations of frustrated magnets predict
that Skyrmions can be tied into knots to form new
stable three-dimensional nanoparticles.
These stable equilibria of twisted loops of Skyrmion strings have
an integer-valued topological charge, known as the Hopf charge, that counts
the number of particles.
Rings are formed for low values of this charge, but for higher values
it is energetically favourable to form links and then knots.
This computational study provides
a novel impetus for future experimental work on these nanoknots and
an exploration of the potential technological applications of three-dimensional nanoparticles encoding knotted magnetization.
\end{abstract}
\maketitle

Lord Kelvin initiated the study of knotted fields by proposing a  
vortex theory of atoms \cite{Kelvin}, where he envisaged vortex strings in the aether forming closed loops that would explain the
distinct atomic species as a simple consequence of the existence of different types of knots and links.
Over the past several years there has been a surge of interest in
knotted fields, with a significant amount of theoretical research in
addition to the physical realization of knots in many diverse systems,
including hydrodynamics \cite{KI}, optics \cite{Den} and liquid crystals \cite{Tka}, to name just a few examples. 
Here we investigate a nanoscale reincarnation of Kelvin's idea and provide numerical support for the existence of stable nanosize knots and links in a particular class of magnetic materials.

  The lead character in our story is the magnetic Skyrmion, a
  two-dimensional topological soliton \cite{book} that lives
  in a thin slice of a magnetic material
  and is characterized by the property that the direction of the
  local magnetization winds exactly once around the sphere of all possible
  directions.
  Magnetic Skyrmions were first obtained experimentally only a few years ago  \cite{Muhl,Yu},  but are now the subject of intense theoretical and experimental research, motivated by technological spintronic applications to information storage and logic devices \cite{Fert}.
  The Skyrmion will play the role of the vortex in our modern
  re-enactment of Kelvin's vision.
  The main idea is to liberate Skyrmions from their two-dimensional world to
  create fully three-dimensional stable topological solitons formed from
  closed loops of twisted Skyrmion strings.

  The majority of current studies on Skyrmions involve a chiral ferromagnetic host. The existence of nanoscale
  Skyrmions in chiral ferromagnets, stabilized by the
  Dzyaloshinskii-Moriya interaction, was predicted by theoretical
  work \cite{BY} in the late 1980's, but it took two decades to obtain
  experimental confirmation \cite{Muhl,Yu}. Chiral ferromagnets are not inversion symmetric, so although they support Skyrmions they do not support anti-Skyrmions (where the magnetization again winds once around the sphere of all possible directions but this time with the opposite orientation).
  An alternative stabilization mechanism is provided by long-range dipolar
  interactions, which allows both Skyrmions and anti-Skyrmions, and these
  have been obtained experimentally \cite{Yu2}
  with a size of around 100\,\mbox{nm}.
  Recent results \cite{Okubo,LM,Lin} on
  Skyrmions in frustrated magnets
  show that these systems are capable of providing the richest
  phenomena. In addition to hosting both Skyrmions and anti-Skyrmions, there
  is an additional rotational degree of freedom that is frozen in other hosts.
  It is this new found flexibility
  that we exploit to obtain
  three-dimensional particles and knots.

  \begin{figure*}[ht]\begin{center}\includegraphics[width=15.0cm]{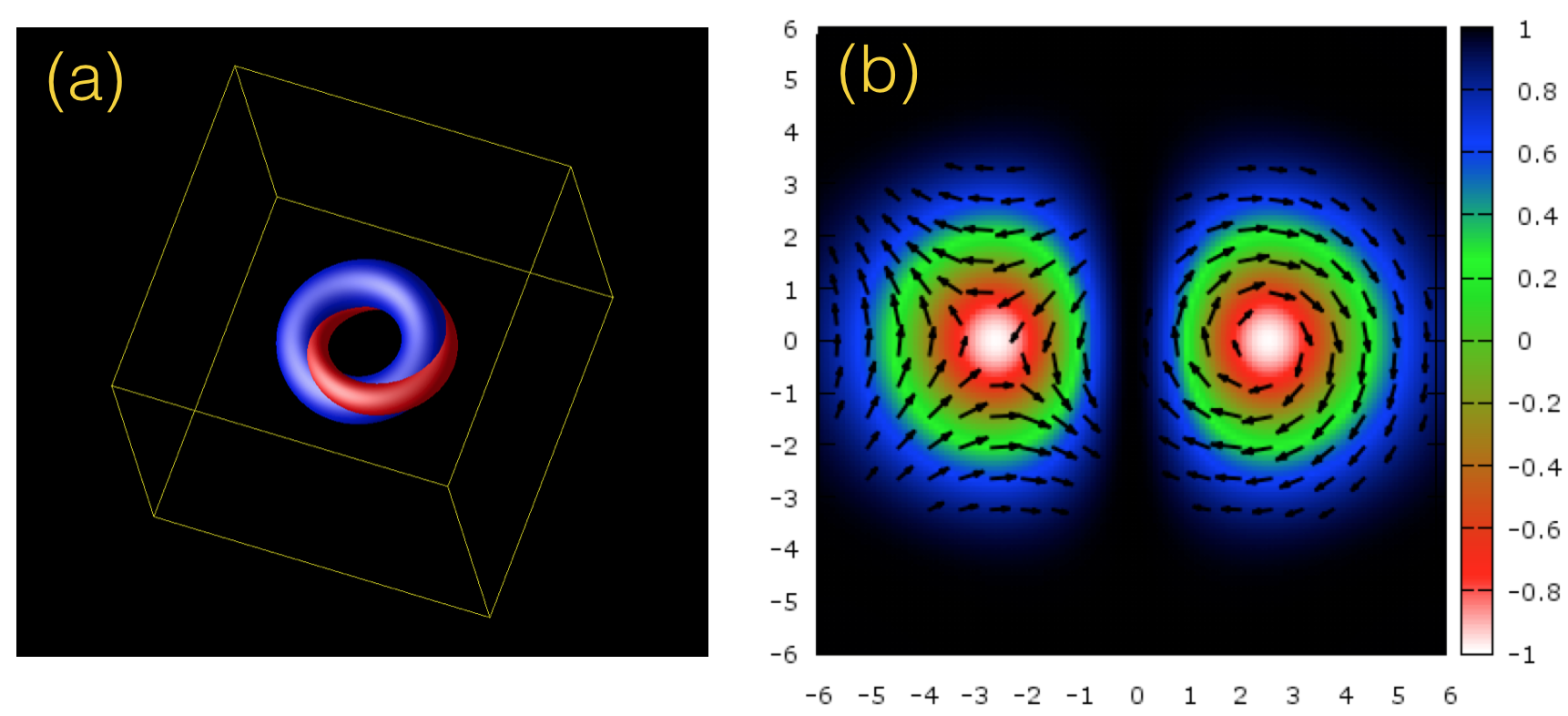}
    \caption{The stable charge one soliton. (a) An isosurface (blue) enclosing the region of the magnetic material where the magnetization points in the direction
      ${\bf m}=(0,0,-1),$ showing the location of the Skyrmion string.
An isosurface (red) enclosing the region  where
${\bf m}=(0,-\frac{1}{\sqrt{2}},-\frac{1}{\sqrt{2}})$ shows the twist
of the Skyrmion string.
This image clearly shows that the linking number, $Q$, is equal to one.
For scale, the outline of the displayed simulation box is a cube of side length 15 dimensionless units.
(b) A cross-section of the magnetization ${\bf m}$ in a plane containing the axis of rotational symmetry. The colour represents the value of $m_3$ and the arrows are proportional to the two-component vector $(m_1,m_2)$. This shows that the right-hand-side of the cross-section contains a Skyrmion and the left-hand-side contains an anti-Skyrmion. Note that the colour schemes in Fig.1a and Fig.1b
are not related because they represent different quantities in the two figures.
    }
    \label{fig-1}  \end{center}\end{figure*}
  We study a frustrated magnet in which there are competing
  nearest-neighbour ferromagnetic and higher-neighbour
  anti-ferromagnetic exchange interactions.
  We employ the continuum Ginzburg-Landau theory for
  these frustrated magnets, which has been
  shown \cite{Lin} to successfully reproduce the
  properties of Skyrmions obtained from lattice computations.
  Within this approach the task is to solve for the spatial distribution
of the direction of the magnetization ${\bf m}({\bf r})$, where ${\bf m}=(m_1,m_2,m_3)$ is a vector of unit length and ${\bf r}=(x,y,z)$ is the spatial coordinate. Only the orientation of the magnetization varies as its magnitude
can be taken to be equal to the constant saturation magnetization.
The static equilibria are found by minimizing
the Ginzburg-Landau energy for the
frustrated magnet, which contains the following leading order terms  \cite{Lin}
\be
E=\int
\bigg(
-\frac{I_1}{2}(\bnabla {\bf m})^2
+\frac{I_2}{2}(\bnabla^2 {\bf m})^2
+|{\bf H}|-{\bf H}\cdot {\bf m})\bigg)\, d^3r,
\label{energy}
\ee
where the constants $I_1$ and $I_2$ are linear combinations of the
nearest-neighbour ferromagnetic and higher-neighbour
antiferromagnetic exchange constants, with the particular linear combinations
depending upon the geometry of the atomic lattice.
The final term is the Zeeman interaction with an external magnetic field ${\bf H},$ which is taken to lie along the third direction ${\bf H}=(0,0,H).$

For Skyrmions in frustrated magnets the crucial feature is that $I_1>0$ and $I_2>0$.
The fact that the first term yields a negative contribution to the energy means that this term favours an increase in scale when considering the standard
three-dimensional Derrick scaling argument, rather than the usual decrease in scale for a ferromagnet with $I_1<0.$
 We use scaled dimensionless units to set $I_1=I_2=1$, so that the nanoscale sized Skyrmion has a size
of order one in dimensionless units.
When the background magnetic field is greater than the saturation field, $H\ge \frac{1}{4}$ in dimensionless units, the ground state is the fully polarized ferromagnetic state ${\bf m}=(0,0,1)$, aligned with ${\bf H}$.
The inclusion of the constant term $|{\bf H}|$ in equation (\ref{energy})
makes use of the freedom to add a constant to the energy and is
included so that the ground state ${\bf m}=(0,0,1)$ is defined to have
zero energy.
We consider isolated solitons embedded in this constant far field by taking the background magnetic field to be twice the saturation value, $H=\frac{1}{2}.$
For simplicity we neglect anisotropy, but in fact anisotropy
improves the stability of Skyrmions in frustrated magnets \cite{LM}, so its
inclusion should only strengthen the results of our findings. 

The energy is computed using a finite difference approximation
on large grids containing $150^3-200^3$ lattice points with a lattice spacing in the
range $0.1-0.4$. Periodic boundary conditions are imposed and
derivatives are approximated by second order accurate finite differences.
The energy is minimized by applying a leapfrog algorithm
to evolve a second order dynamics in an auxiliary time with periodic
removal of auxiliary kinetic energy \cite{Su}.
We have verified that the results are stable to changes in the number of
grid points and the lattice spacing.

The three-dimensional topology is provided by the Hopf charge \cite{Atiyah}, $Q$, which is an integer that counts the linking number of magnetization field lines (this being the appropriate topological generalization of the Skyrmion winding number in two dimensions). In detail, consider all the points in the material where the
direction of the magnetization is equal to ${\bf m}=(0,0,-1)$, which is
exactly the opposite direction to the far field magnetization. 
In a three-dimensional medium this value of
the magnetization will generically be attained on a closed curve in space (or
potentially a collection of closed curves). This closed curve specifies the location of the Skyrmion string, as one of the characteristic features of the
Skyrmion is that the magnetization at its centre points in exactly
the opposite direction to that of the far field. Now consider any other
fixed direction for the magnetization,
a convenient choice for visualization is to take
${\bf m}=(0,-\frac{1}{\sqrt{2}},-\frac{1}{\sqrt{2}}),$
but any other direction will do just as well to compute the topology.
This second direction will define a second closed curve in the material by the same reasoning. The Hopf charge is the integer that counts the number of times these two closed curves are linked with each other. It is a mathematical theorem that this Hopf charge is topological (it does not change under any smooth deformation of the field ${\bf m}({\bf r})$) and that its value does not depend upon
the choice of the two particular fixed directions used to define the two closed curves. In other words, if two new directions are chosen then this will yield two new curves but the linking number will be unchanged.

A simple explicit example of a magnetization field with $Q=1$ is given by
\be
   {\bf m}({\bf r})=\frac{4}{(\lambda^2+r^2)^2}
   \begin{pmatrix}2\lambda^2xz-\lambda y(\lambda^2-r^2)\\
     2\lambda^2yz+\lambda x(\lambda^2-r^2)\\
     2\lambda^2z^2-\lambda^2r^2+\frac{1}{4}(\lambda^2-r^2)^2
   \end{pmatrix}
   \ee
   and has the topology of a circular Skyrmion string, with radius $\lambda$,
   lying in the plane $z=0$ and twisted once.
   The above expression can be used as an initial field for a
   relaxation of the energy. The resulting stable equilibrium
   magnetization is displayed in Fig.1. This ring-like charge one soliton
   consists of a closed circular Skyrmion string, where the Skyrmion is twisted once so that the first two components of the magnetization perform one
   full revolution as the circle is traversed. The characteristic winding of
   the Skyrmion is seen in cross-section in the right-half of the image
   in Fig.1b, where the Skyrmion string moves into the plane,
   and again in the left-half of the image, where the Skyrmion string
   returns to the plane, but this time moving out of the plane
   so that it appears as
   an anti-Skyrmion in cross-section.
   Applying random perturbations confirms that this static three-dimensional
   soliton solution with charge $Q=1$ is stable. In our dimensionless units,
   the energy of this charge one solution is computed to be $E_1=280,$
   where this is the energy above the zero energy ground state given by
the fully polarized ferromagnetic state ${\bf m}=(0,0,1)$, aligned with ${\bf H}$. 
   
   \begin{figure}[ht]\begin{center}\includegraphics[width=8.0cm]{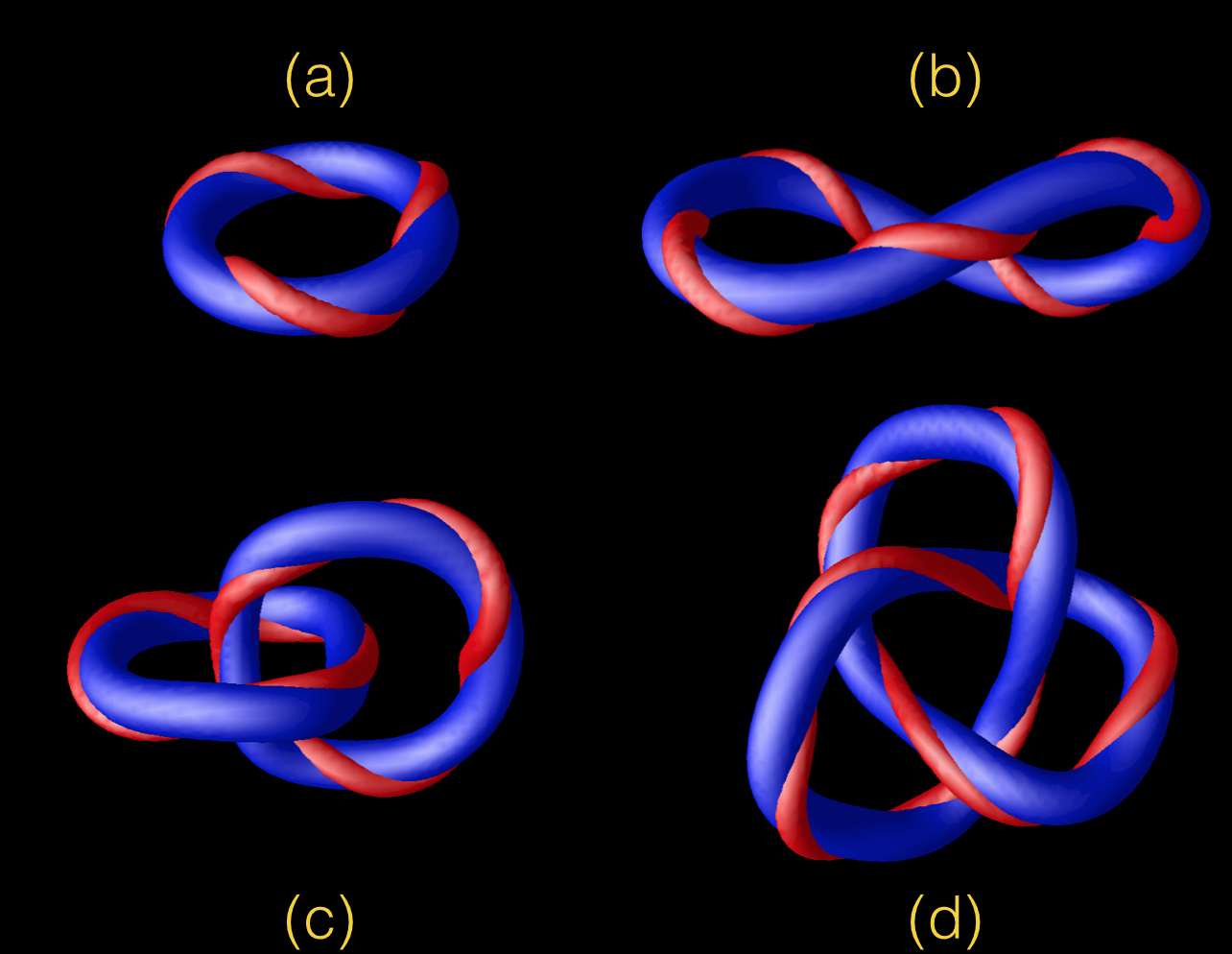}
    \caption{Stable higher charge solitons. Isosurfaces enclosing the region of the magnetic material
where the magnetization points in the directions ${\bf m}=(0,0,-1)$
      (blue) and 
     ${\bf m}=(0,-\frac{1}{\sqrt{2}},-\frac{1}{\sqrt{2}})$
      (red), for a selection of stable solitons: (a) the charge 3 circular ring, (b) the charge 6 buckled ring, (c) the charge 7 link with components of charge 3 and 2, (d) the charge 10 trefoil knot.
    }
    \label{fig-36710}  \end{center}\end{figure}
   
To study higher charge solutions ($Q>1$) requires initial fields
with any prescribed value of $Q$, as starting configurations for the
energy minimizing relaxation algorithm.
Fortunately, there is a simple method \cite{Su} that uses complex
polynomials to create initial rings with any value of $Q$, corresponding to
a Skyrmion string twisted $Q$ times. Furthermore, this method can also
generate initial fields describing twisted Skyrmion strings that are
knotted or linked, by using mathematical representations of knots and links in terms of the intersection of complex curves with the three-sphere.
All initial fields created using this method are first
perturbed by a stretch along a random direction to break any potential symmetries of the initial field, to ensure a thorough test of the stability properties
of the resulting static equilibria.

The results of the energy minimization computations reveal that stable ring
equilibria, analogous to the $Q=1$ ring, exist 
 for higher values of the Hopf charge $Q$,
   with the Skyrmion string twisted $Q$ times (the $Q=3$ example is presented in Fig.2a). However, for $Q>4$ the circular ring is unstable to a buckling instability that results in a stable buckled ring with lower energy
   (see Fig.2b for the $Q=6$ example).
   The energies per unit Hopf charge $E/Q$, in units of the charge one energy $E_1$,  are displayed as the black circles in Fig.3 for these circular or buckled rings. For clarity, the $Q=2$ energy, computed to be $E/Q=0.73E_1$,
   is not shown on this plot.
   
   The description of the topological Hopf charge in terms of a linking number
   reveals that a ring with charge $Q_1$ linked once with a ring with charge $Q_2$ yields a combined configuration with total Hopf charge equal to $Q=Q_1+Q_2+2$, where the extra $+2$ is due to the linking of each ring once with the other. Initial fields of this type revert to stable rings under minimization of the energy for $Q\le 6$ but for $Q=7$ there is a stable link equilibrium with $Q_1=3$ and $Q_2=2$, presented in Fig.2c. Computations reveal that stable links exist for $Q\ge 7$ and the energy is lowest if the charge is distributed as equally as possible between the two components, so that $Q_1=Q_2=(Q-2)/2$ if $Q$ is even and $Q_1=Q_2+1=(Q-1)/2$ if $Q$ is odd. Furthermore, for $Q\ge 8$ these stable links have lower energy than the corresponding buckled rings. The energies of these links
   are displayed in Fig.3 as the red squares.

   If a Skyrmion string is twisted $Q_1$ times and tied into a trefoil knot
   before joining the ends then this yields a charge $Q=Q_1+3$, where the
   $+3$ results from the crossing number of the trefoil knot (the simplest non-trivial knot). Under minimization of the  energy, initial fields of this type revert to stable rings for $Q<7$ and to stable links for $Q=7,8,9$.
   Remarkably, for $Q\ge 10$ stable trefoil knot equilibria are obtained with energies below the corresponding rings. Moreover, for $Q\ge 12$
   these knot energies are also below those of the corresponding links.
   The knot energies are displayed in Fig.3 as the blue triangles.
   The first stable trefoil knot is shown in Fig.2d, where 7 twists can be seen to contribute to the $Q=10$ charge.
   A comparison of the energies of rings, links and knots is presented in Fig.3  and confirms the general conclusion that rings are formed for low values of the
   Hopf charge, but for higher values it is energetically favourable to form links and then knots. 
   
\begin{figure}[ht]\begin{center}\includegraphics[width=8.0cm]{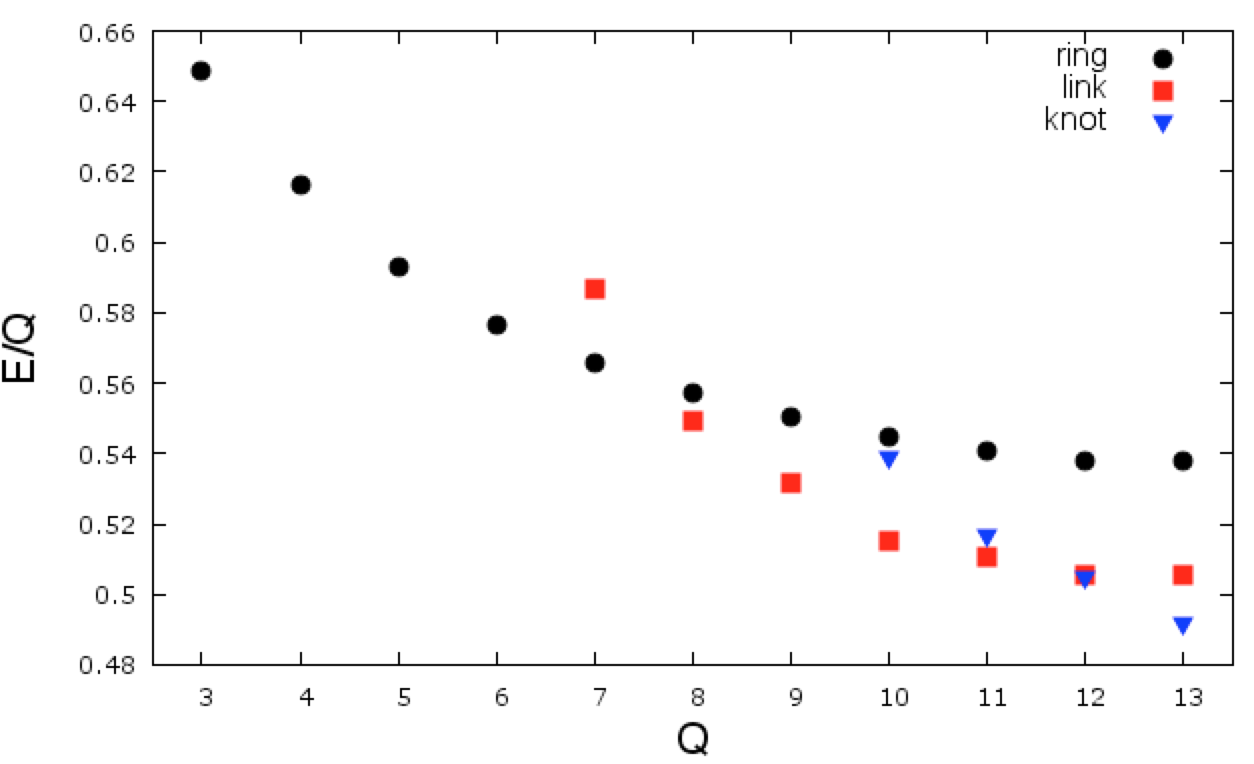}
    \caption{Soliton energies. The energy per unit charge $E/Q$, in units of the charge one energy, for stable equilibria that take the form of circular or buckled rings (black circles), links (red squares) and knots (blue triangles).  
    }
    \label{fig-energy}  \end{center}\end{figure}

   Finally,
   we place our results in the context of previous work on Hopf solitons.
   The solitons presented here have a number of similarities to those
   in the mathematical Skyrme-Faddeev model, where stable knotted solitons
   were first conjectured to exist \cite{FN} and subsequently computed
   numerically \cite{BS,HS}. Note that this is despite the fact that
   the term quadratic in derivatives has opposite sign in the two systems.
   Dynamical Hopf soliton rings exist \cite{Cooper,Su2}
   as solutions of the Landau-Lifshitz
   equations describing the evolution of the magnetization in a
   ferromagnetic material. These circular magnetic smoke rings drift
   along their symmetry axis but because they are
   dynamical they decay due to dissipation, unlike the static solitons
   considered here. Furthermore, the dynamical stabilization mechanism
   favours rings and not links or knots.

   Several possible candidate materials have been suggested \cite{Okubo,LM} as
   frustrated magnetic hosts for Skyrmions and significant experimental investigation is required to turn these possibilities into reality.
   The fact that the solitons are spatially localized in all directions
   within a constant far field ground state means that they are robust
   to changes in the system size and do not require the imposition
   of particular boundary conditions.
   It will certainly be a challenging task to engineer these complicated
   knotted structures but there are a number of recent exciting
   proposals for Skyrmions that could be adapted to this context.
   Examples include imprinting spatial structure in the magnetization
   field using optical vortex beams \cite{FuSa}, using nanostructured
   templates formed from arrays of ferromagnetic nanorods \cite{Del}, or
   seeding topological structures via nonmagnetic impurities \cite{LHB}.
   Advanced magnetic imaging techniques for Skyrmions are rapidly developing,
   so the plethora of microscopy methods being investigated
   could allow the identification of the twist in a Skyrmion string
   that is the characteristic signature of these solitons.

   In summary, the computational results presented in this letter predict a zoo of exotic rings, links and knots waiting to be discovered in the nanoscale world of frustrated magnets.
   We hope that the tantalizing prospect of a nanoscale
  resurrection of Kelvin's dream of knotted fields will provide a novel
  impetus for future work, together with an exploration of the
  potential technological applications of  three-dimensional
nanoparticles encoding knotted magnetization.\\

  {\bf \small \qquad \qquad ACKNOWLEDGEMENTS}
  
\noindent  
This work is funded by the
Leverhulme Trust Research Programme Grant RP2013-K-009, SPOCK: Scientific Properties Of Complex Knots, and the STFC grant ST/J000426/1.

\end{document}